\documentclass{article}

\usepackage[utf8]{inputenc}
\usepackage{amsmath, amsthm, amssymb, amsfonts, dsfont, mathrsfs, dirtytalk, tikz-cd, adjustbox, url, nicefrac, booktabs,array, graphicx,todonotes,longtable, tabu,bm, todonotes,comment,multirow,tabularx,epsfig,parskip}
\usepackage[T1]{fontenc}    
\usepackage[margin=1in]{geometry}
\usepackage[english]{babel}
\usepackage{hyperref}
\usepackage[capitalise]{cleveref}
\usepackage[backend=biber,style=numeric,isbn=false,url=false]{biblatex}
\newcommand{\f}{\forall}

\newcommand{\dkl}{\operatorname{D_{K L}}}
\renewcommand{\d}{\: \mathrm{ d }}

\theoremstyle{plain}
\newtheorem{theorem}{Theorem}[section]

\theoremstyle{definition}

\theoremstyle{remark}
\newtheorem{remark}[theorem]{Remark}

\newcounter{daggerfootnote}

\title{Metacognitive particles, mental action and the sense of agency}
\author{Lars Sandved-Smith$^{1,}$\thanks{Equal contribution. Correspondence: \url{lars.sandvedsmith@gmail.com}}\: , Lancelot Da Costa$^{2,3,4,*}$\\\\
$^1$\textit{Monash Centre for Consciousness and Contemplative Studies, Monash University, Australia}\\
$^2$\textit{VERSES AI Research Lab}\\
$^3$\textit{Department of Mathematics, Imperial College London} \\
$^4$\textit{Wellcome Centre for Human Neuroimaging, University College London}\\
}
\date{\vspace{-10pt}}

\begin{document}
\maketitle
\begin{abstract}
This paper articulates metacognition using the language of statistical physics and Bayesian mechanics. Metacognitive beliefs, defined as beliefs \emph{about beliefs}, find a natural description within this formalism, which allows us to define the dynamics of `metacognitive particles', i.e., systems possessing metacognitive beliefs. We further unpack this typology of metacognitive systems by distinguishing \emph{passive} and \emph{active} metacognitive particles, where active particles are endowed with the capacity for mental actions that update the parameters of other beliefs. We provide arguments for the necessity of this architecture in the emergence of a subjective sense of agency and the experience of being separate from the environment. The motivation is to pave the way towards a mathematical and physical understanding of cognition -- and higher forms thereof -- furthering the study and formalization of cognitive science in the language of mathematical physics.
\end{abstract}


\tableofcontents

\section{Introduction}

In this theoretical paper we provide a link between classical physics and metacognition. We ask: what could metacognition look like in simple physical terms? We adopt a Bayesian mechanical lens, where we define metacognition as having beliefs about beliefs. Under this definition, metacognition can be articulated simply using the language of statistical physics. We explore how the resulting architecture gives rise to formal notions of mental action and a subjective sense of agency. We then discuss connections to computational phenomenology, making preliminary empirical predictions and highlighting directions for future work. The motivation is to pave the way towards a mathematical and physical understanding of cognition -- and its higher forms -- furthering the study and formalization of cognitive science using mathematical physics.

\section{Systems, states and fluctuations}


We consider a system over some period of time. For simplicity, and ease of exposition, we assume that the system evolves according to a stochastic differential equation (a.k.a. Langevin equation):

\begin{equation}
\label{eq: Langevin}
    \dot x(t)= f(x(t))+w(t).
\end{equation}

This equation decomposes the motion of the system over some state space in terms of what we know about the system, specified in terms of the \emph{flow} $f$ -- a vector field on the state space -- and what we don't know about the system; summarized by a noise process $w$, that represents \emph{random fluctuations}, and which is usually assumed to be a mean-zero stationary Gaussian process (by the central limit theorem). This functional form for the dynamics will conveniently enable us to write down the causal relationships between different subsets of states in terms of the flow later on.

This is a natural place to start because much of physics, and in particular statistical physics, quantum mechanics and classical mechanics can be formulated with stochastic differential equations. In short, if we want a physics of cognition that is compatible with the rest of physics, then this is the right place to start.

A stochastic differential equation is an implicit specification of the random trajectories of the system, as a function of the flow and the random fluctuations. The fact that the fluctuations are random means that the system also exhibits a degree of randomness, and can only be described probabilistically.

\subsection{Notation}

In what follows, we denote trajectories of the system by $x$ (a random variable on a space of paths\footnote{We will refer to ‘paths’ as opposed to ‘states’ for technical reasons (that inherit from the path integral formulation of the free energy principle). Paths can be thought of as trajectories or events that traverse state space and behave, mathematically, very much like a variable or state. Indeed, a path can be treated as a state in generalised coordinates of motion.} over state space) and trajectories of the noise process by $w$ (idem). We will denote the state of the system, and the state of random fluctuations at some time point $t$ by $x(t)$ and $w(t)$ (random variables over state space). Each of these random variables can be equivalently described by a probability distribution, over state space or path space, e.g. $P(x(t))$ and $P(x)$ respectively. We use capital $P$ to emphasize that we operate with distributions over random variables which may or may not admit densities. For intuition, it is useful to think of them as densities though. We will use an analogous notation for all other stochastic processes that we will encounter later, i.e. subparts of the system.

\section{Particles and things}

\begin{figure}[h]
\centering\includegraphics[width=\textwidth]{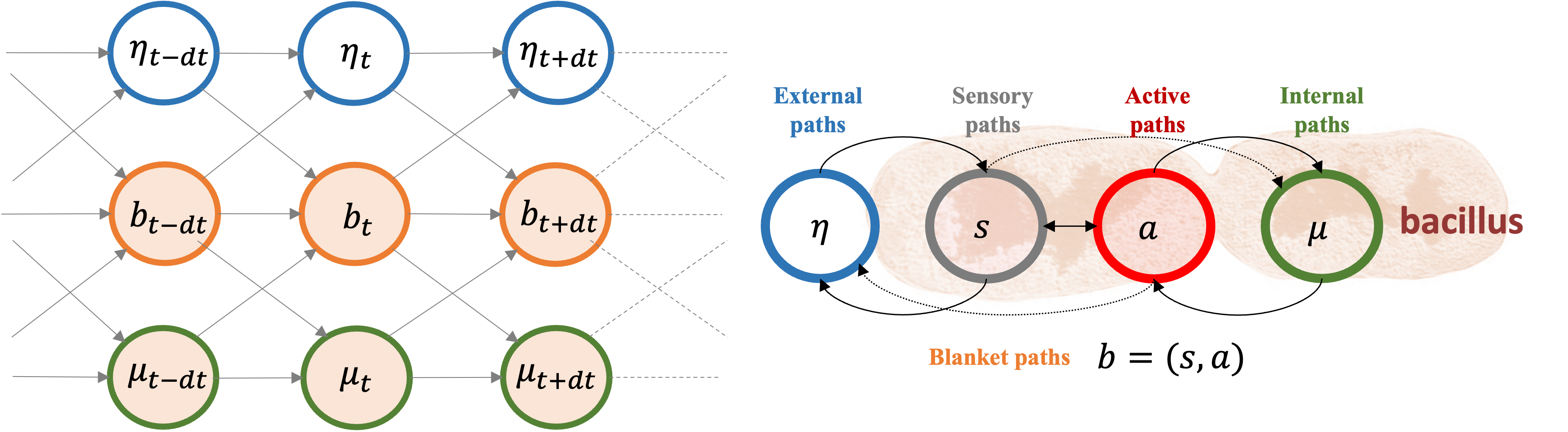}
\caption{\textbf{Particles and Markov blankets.} The left diagram illustrates the temporal unfolding of external paths (blue) and internal paths (green), as separated by blanket paths (orange). Internal and external paths can only interact via the intermediary blanket paths. The right diagram depicts the conditional dependencies of the paths of a system $x=(\eta, b, \mu)$. Arrows represent the direction of influence between paths. A key aspect of this interaction is that the external paths can only impact themselves and sensory paths, while the internal states are restricted to influencing themselves and the active paths. By increasing the sparsity and removing the interaction from active paths to internal paths we recapture the notion of a \emph{strange} particle defined in \cite{fristonPathIntegralsParticular2023}.}
\label{fig: particle}
\end{figure}

Any object of study in our system -- over the period of time that it is observed -- must by definition be distinguishable from the rest of the system. This in turn implies a boundary that separates the internal paths of the object from its external paths. We thus partition the system into external, boundary and internal:

\begin{equation}
    x=(\eta,b,\mu).
\end{equation}

Mathematically, the boundary paths must constitute a \textit{Markov blanket} between the external and internal paths

\begin{equation}
 \label{eq: def Markov blanket}  
          \eta \perp \mu \mid b
 \iff P(\eta, \mu | b) = P(\eta | b)P(\mu | b)
  \iff  P(\eta | b, \mu) = P(\eta | b)   \iff P(\mu | b, \eta) = P(\mu | b).
\end{equation}

\eqref{eq: def Markov blanket}  formalises the physical intuition that all interactions between internal and external states happen via the boundary. See Figure \ref{fig: particle} (left) for an illustration. 

For this reason, we will use the terms boundary and blanket interchangeably henceforth. In turn, we will call the object of study, thing, or person – formed by internal and blanket states – a particle. This is in reference to the fact that a particle could describe a simple microscopic particle from statistical physics, a large particle like a planet as considered in classical mechanics or general relativity, or a biological organism  – like a cell or a human being – as considered in biophysics. See Figure \ref{fig: particle} (right).

Perhaps the simplest functional form for the dynamics that guarantees a Markov blanket is as follows

\begin{equation}
     \begin{bmatrix}\dot \eta \\ \dot b \\\dot \mu \end{bmatrix}(t)=
     \begin{bmatrix}f_\eta(\eta, b) \\ f_b(\eta, b, \mu) \\f_{\mu}(b, \mu)\end{bmatrix}(t)+
     \begin{bmatrix}w_\eta\\ w_b\\w_{\mu}\end{bmatrix}(t),
\end{equation}
when the random fluctuations on external, blankets and internal states are independent.

\subsection{Sensorimotor boundaries}

In turn, we subdivide the boundary or blanket into what we call sensory and active paths

\begin{equation}
    b=(s,a).
\end{equation}

We operationally define sensory paths as those boundary paths that influence internal paths directly, but are not directly influenced by internal paths. In turn, we define active paths as those which influence external paths directly but are not directly influenced by external paths; then the motion of the system may read as follows:

\begin{equation}
     \begin{bmatrix}\dot \eta \\\dot s\\\dot a \\\dot \mu \end{bmatrix}(t)=
     \begin{bmatrix}f_\eta(\eta, s, a) \\ f_s(\eta, s, a) \\ f_a(s, a, \mu)\\f_{\mu}(s, a, \mu)\end{bmatrix}(t)+
     \begin{bmatrix}w_\eta\\ w_s\\w_a\\w_{\mu}\end{bmatrix}(t).
\end{equation}

See Figure \ref{fig: particle} (right) for an illustration.


\subsection{Cognitive and metacognitive particles} 

We say that a particle is \emph{cognitive} whenever its internal paths parameterise beliefs about its external paths. Belief here is used in a technical sense to mean a conditional (i.e., Bayesian) probability distribution parameterised by some sufficient statistics. The defining property here is that there exists an assignment from internal paths to beliefs about external paths such that the most likely internal path given blanket paths encodes the posterior belief about external paths:\footnote{With an abuse of notation for taking the maximum of a distribution; see \cite{durrOnsagerMachlupFunctionLagrangian1978} for the rigorous definition of this in our setting.}
\begin{equation}
\begin{split}
\mu &\mapsto Q_{\mu}(\eta)\\
    Q_{\boldsymbol{\mu}}(\eta) & \triangleq P(\eta \mid s, a) \\
\boldsymbol{\mu} & \triangleq \arg \max _{\mu} P(\mu \mid s, a)
\end{split}
\end{equation}
Intuitively, this means that internal paths track the external world given the information on the boundary. Sufficient conditions for a particle to be cognitive are given in \cite{fristonPathIntegralsParticular2023,dacostaBayesianMechanicsStationary2021a,parrMarkovBlanketsInformation2020}. Note that in some cases (i.e. under the generalised coordinate formulation of a stochastic differential equation) paths are parameterised by states, so that internal states parameterise beliefs about external paths \cite{fristonPathIntegralsParticular2023}; but we will not delve into these technicalities here.

\begin{remark}
In existing treatments of the free-energy principle, particles are taken to be cognitive by definition \cite{fristonPathIntegralsParticular2023,fristonFreeEnergyPrinciple2019a,fristonFreeEnergyPrinciple2023a}. We deliberately make a distinction here by saying that a particle is defined in terms of a Markov blanket that exists over some period of time -- which corresponds to the definition of what it is to exist -- while a cognitive particle is defined as a particle with approximate posterior beliefs (and implicitly internal states that can be individuated from boundary states in the well-defined sense above).
\end{remark}

A \emph{metacognitive} particle is a cognitive particle that has beliefs about its own beliefs (about external states of the world). This means that a subset of the internal states, say $\mu^{(2)}$, encodes posterior beliefs about another subset of the internal states, say $\mu^{(1)}$. We will call $\mu^{(2)}$ the \textit{higher}-level internal paths and $\mu^{(1)}$ the \textit{lower}-level internal paths in virtue of the fact that $\mu^{(2)}$ will be seen as encoding beliefs about $\mu^{(1)}$.

We make a further the distinction between \emph{passive} metacognition and \emph{active} metacognition. Passive metacognitive beliefs are parameterised by a subset of internal paths such that they can only influence the lower-level beliefs via their shared blanket paths (see Figure \ref{fig: metacognitive particle}). For example, a person who is surprised by their sudden inability to taste coffee (a possible sign of pancreatic cancer) might introspect and act accordingly \cite{hohwyThePredictiveMind2013}. Active metacognitive beliefs are parameterised by higher-level internal paths that are separated from the lower-level paths by an internal Markov blanket (see Figure \ref{fig: nested particle}). The term \textit{active} refers to the existence of higher-level active paths that influence the lower-level internal paths (see Section \ref{sec: active metacognition}). 

Note that in the active case, $\mu^{(1)}$ can constitute the \emph{entire set of parameters} of the particle's belief about the world. In contrast, passive metacognitive beliefs are necessarily about \emph{a subset of the parameters} of the particle's belief about the world.

We now present some examples to see what this looks like.

\section{An example of a passive metacognitive particle}
\label{sec: passive metacognition}

\begin{figure}[h]
\centering\includegraphics[width=0.7\textwidth]{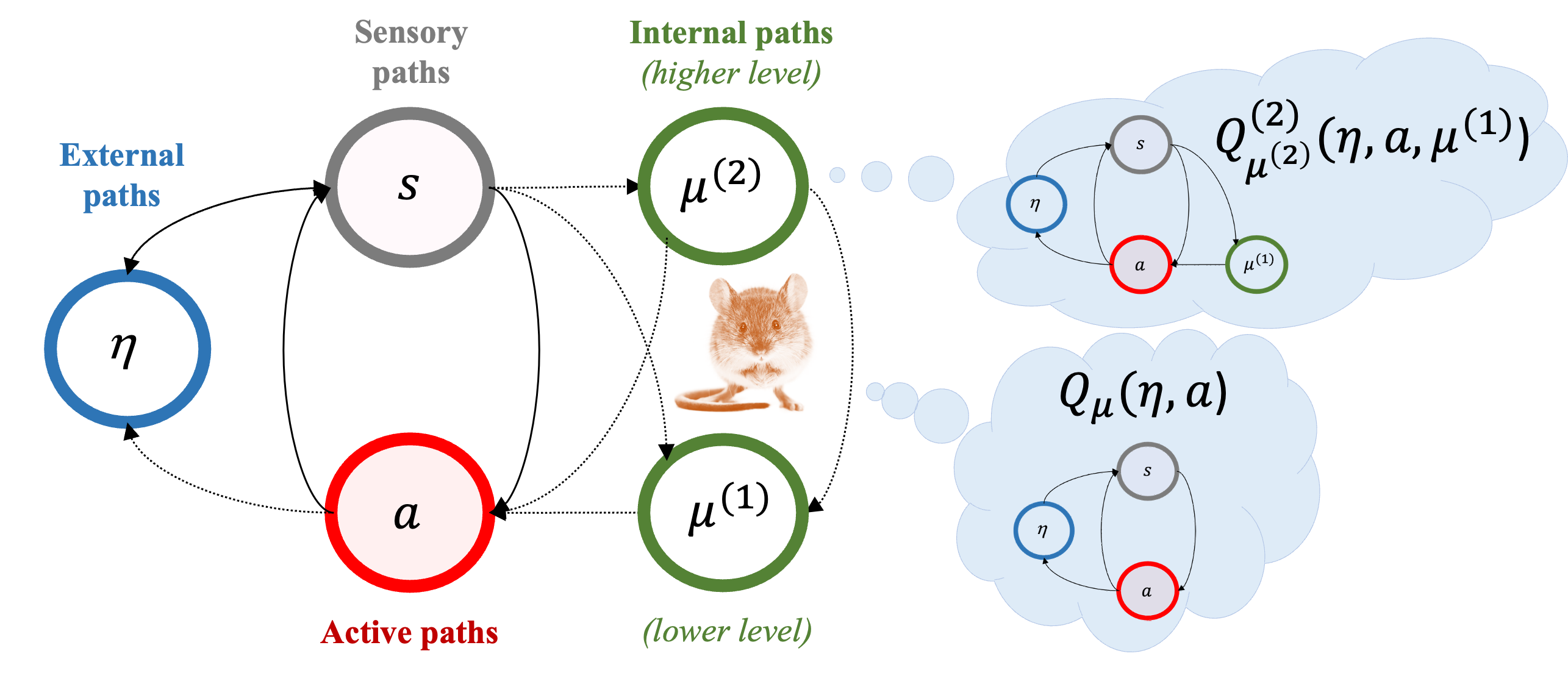}
\caption{\textbf{Example of a passive metacognitive particle.} This particle is such that internal paths at the lower level (i.e. $\mu^{(1)}$) are hidden from internal paths at the higher level (i.e. $\mu^{(2)}$). In other words, higher level internal paths must infer the lower level internal paths via the blanket paths, leading to the metacognitive belief $Q_{\mu^{(2)}}^{(2)}(\mu^{(1)})$. Because the lower level internal paths $\mu^{(1)}$ constitute only a subset of the parameters $\mu=(\mu^{(1)},\mu^{(2)})$ of the particle's belief about the world, and there are no higher level active paths; the particle has \emph{passive} metacognition.}
\label{fig: metacognitive particle}
\end{figure}

We rehearse an example of a passive metacognitive particle, first proposed in \cite{dacostaBayesianMechanicsMetacognitive2024}.

Consider a strange particle\footnote{Strange particles, as defined in \cite{fristonPathIntegralsParticular2023}, are such that active paths do not directly influence (i.e. are hidden from) internal paths; and, accordingly, active paths are inferred by internal paths via sensory paths. The other defining condition is that random fluctuations on the blanket and internal states are negligible in virtue of the fact that the particle is assumed to be large, and modelled at a correspondingly large degree of coarse graining (cf. classical mechanics from statistical mechanics).} whose internal dynamics can be decomposed into two sets of paths $\mu\triangleq(\mu^{(1)},\mu^{(2)})$, and such that the first set of paths is hidden from the second via the Markov blanket. That is, $\mu^{(2)}$ may influence $\mu^{(1)}$ directly, but $\mu^{(1)}$ may only influence $\mu^{(2)}$ vicariously, via the Markov blanket. To visualize and summarise, the defining system's dynamics are as follows:
\begin{equation}
\label{eq: flow}
    \begin{split}
        \begin{bmatrix}\dot{\eta} \\ \dot{s} \\ \dot{a} \\ \dot{\mu}^{(1)}\\\dot{\mu}^{(2)} \end{bmatrix}(t)=  \begin{bmatrix}f_\eta(\eta, s, a) \\ f_s(\eta, s, a) \\ f_a(s, a, \mu) \\ f_{\mu^{(1)}}(s, \mu)\\f_{\mu^{(2)}}(s, \mu^{(2)})\end{bmatrix}(t)+\begin{bmatrix}w_\eta \\ 0 \\ 0 \\ 0\\0\end{bmatrix}(t)
    \end{split}
\end{equation}
The form of the coupling can also be seen in Figure \ref{fig: metacognitive particle}. 

In this setting, it can be shown that sensory paths form a Markov blanket between higher- and lower-level internal paths \cite[eq. 29]{fristonPathIntegralsParticular2023}. Put simply, given the sensory paths, there is no further information in the lower level paths that is not inherent in the higher level paths.
\begin{equation}
(\mu^{(2)} \perp \eta, a, \mu^{(1)}) \mid s
\end{equation}
In particular, the lower-level internal paths can only be inferred vicariously by the higher-level internal paths via the sensory paths. So we can define \textit{metacognitive} beliefs encoded by the higher-level internal paths\footnote{These exist in large class of particles in virtue of conditions analogous to that discussed in \cite[p. 24]{fristonPathIntegralsParticular2023}.} 
\begin{equation}
\label{eq: meta}
\begin{split}
\mu^{(2)}&\mapsto Q_{\mu^{(2)}}^{(2)}(\eta, a,\mu^{(1)})\\
Q_{\boldsymbol{\mu}^{(2)}}^{(2)}(\eta, a,\mu^{(1)})  &\triangleq P(\eta, a ,\mu^{(1)}\mid s) \\
\boldsymbol{\mu}^{(2)}  &\triangleq \arg \max _{\mu^{(2)}} P(\mu^{(2)} \mid s)
\end{split}
\end{equation}
This sort of belief is interesting because it implies that the higher internal paths encode beliefs about beliefs, licensing the metacognitive terminology. Perhaps the easiest way to see this is when the particular beliefs factorise according to a mean-field approximation:
\begin{equation}
    Q_{\mu}(\eta)=Q_{\mu^{(2)}}(\eta)Q_{\mu^{(1)}}(\eta)
\end{equation}
Then the (marginal) metacognitive belief 
\begin{equation}
    Q_{\mu^{(2)}}^{(2)}(\mu^{(1)}) \triangleq \iint Q_{{\mu}^{(2)}}^{(2)}(\eta, a,\mu^{(1)}) \d \eta \d a
\end{equation}
is a belief about the belief $Q_{\mu^{(1)}}(\eta)$. The upshot is that this particle has \emph{passive} metacognition because it has beliefs \textit{only} about a subset of the parameters of its beliefs about the world, with no ability to influence these directly. See \cite{hespDeeplyFeltAffect2021} for an example of an active inference simulation of a passive metacognitive particle.

The reader may notice that metacognitive beliefs are about the parameters of lower level beliefs, and wonder whether this justifiably entails \emph{beliefs about beliefs}. We point out that indeed it does: a Bayesian belief is a probability distribution, which may be fully described -- as is the case here -- by its parameters or sufficient statistics. Defining a probability distribution over the sufficient statistics, or parameters, of a belief thus amounts to a (metacognitive) belief about this belief.

Next we consider a different class of particles that is comprehensively metacognitive in the richer sense that it has beliefs about all of the [the sufficient statistics of] its beliefs about the world. These are nested particles.

\section{Nested particles and active metacognition}
\label{sec: active metacognition}

\begin{figure}[h]
\centering\includegraphics[width=\textwidth]{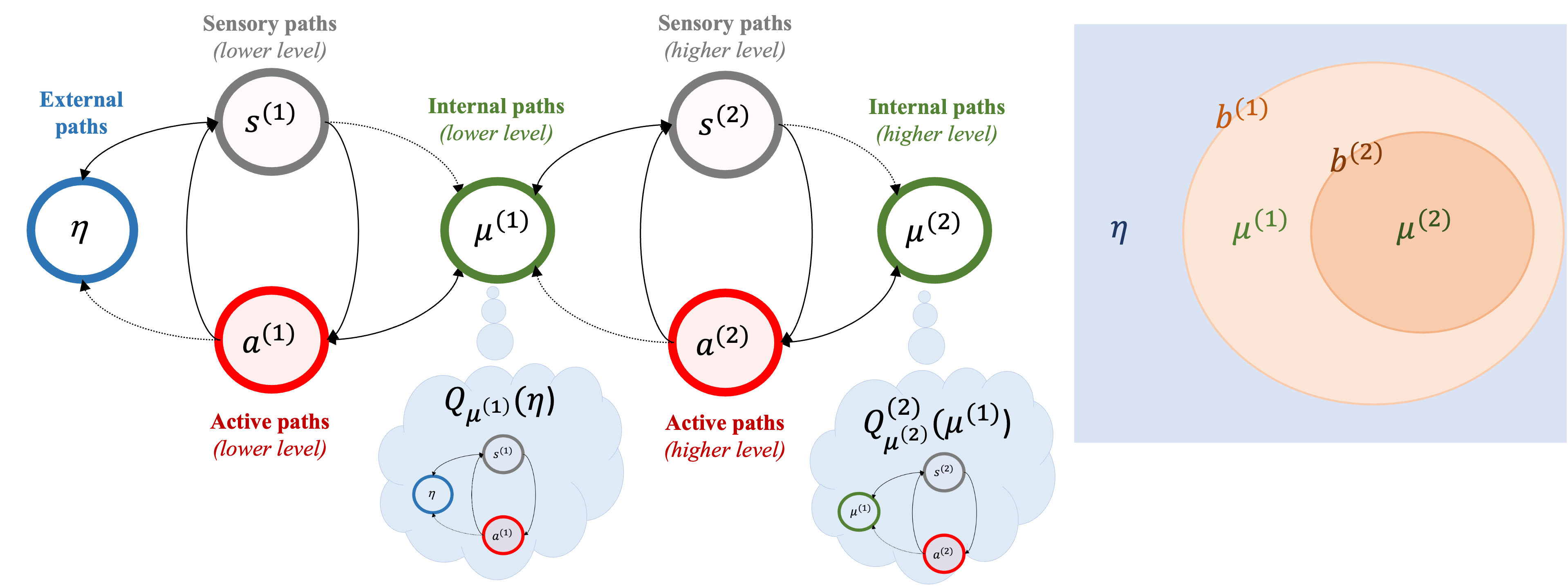}
\caption{\textbf{Nested particles.} This diagram illustrates the decomposition of a system $x$ that contains a nested metacognitive particle. The right diagram shows the nested structure of a metacognitive particle, which possesses an internal Markov blanket separating higher-level internal paths $\mu^{(2)}$ from lower-level internal paths $\mu^{(1)}$. The left diagram depicts the conditional dependencies between paths of a metacognitive system. Thought bubbles illustrate the Bayesian beliefs parameterised by the internal paths, about paths external to them.}
\label{fig: nested particle}
\end{figure}

Here we consider nested particles, i.e. particles within particles. See Figure \ref{fig: nested particle} for an illustration. We defer the dynamics of a nested particle to Appendix \ref{app: flow}
 as it is not essential for our discussion. The point is that the lower level internal paths $\mu^{(1)}$ (from the outer particle) can encode posterior beliefs about external paths, given lower level blanket paths:
\begin{equation}
\label{eq 1}
\begin{split}
\mu^{(1)} &\mapsto Q_{\mu^{(1)}}(\eta)\\
    Q_{\boldsymbol{\mu}^{(1)}}(\eta) & \triangleq P(\eta \mid s^{(1)}, a^{(1)}) \\
\boldsymbol{\mu}^{(1)} & \triangleq \arg \max _{\mu^{(1)}} P(\mu^{(1)} \mid s^{(1)}, a^{(1)})
\end{split}
\end{equation}
Further, the internal paths of the inner particle can hold beliefs about the lower level internal paths, which from the perspective of the inner particle are external paths: 

\begin{equation}
\label{eq 2}
\begin{split}
\mu^{(2)}&\mapsto Q_{\boldsymbol{\mu}^{(2)}}(\mu^{(1)})\\
    Q_{\boldsymbol{\mu}^{(2)}}(\mu^{(1)}) & \triangleq P( \mu^{(1)}\mid s^{(2)}, a^{(2)}) \\
\boldsymbol{\mu}^{(2)} & \triangleq \arg \max _{\mu^{(2)}} P(\mu^{(2)} \mid s^{(2)}, a^{(2)})
\end{split}
\end{equation}

In is the case, the higher level internal paths $\mu^{(2)}$ are metacognitive, in the sense that they parameterise beliefs about the parameters $\mu^{(1)}$ of beliefs about the world. This form of hierarchical depth can therefore be described as a `parametric depth' \cite{sandved-smithComputationalPhenomenologyMental2021}. For instance, if beliefs about the world are Gaussian:

\begin{equation}
    Q_{{\mu}^{(1)}}(\eta)=\mathcal N(\eta ; m, \Pi)
\end{equation}
where the mean $m= m({\mu}^{(1)})$ and the precision $\Pi=\Pi({\mu}^{(1)})$ are functions of the sufficient statistic ${\mu}^{(1)}$.
Then higher level internal paths $\mu^{(2)}$ possess beliefs about the sufficient statistics of that Gaussian. The higher level internal paths $\mu^{(2)}$ sense those parameters via the higher level sensory paths $s^{(2)}$ that are constituted by the higher level boundary. 

In turn, the active trajectories at the higher level $a^{(2)}$ can modulate those parameters of beliefs about the world (e.g. posterior precision, which can be seen as a proxy for attention \cite{sandved-smithComputationalPhenomenologyMental2021}). This formally encapsulates notions of `mental actions' that modulate the parameters of a lower level generative model (cf. \cite{sethInteroceptiveInferenceEmotion2013,clarkManyFacesPrecision2013,limanowskiDisAttendingBody2017,limanowskiSeeingDarkGrounding2018} and \cite{sandved-smithComputationalPhenomenologyMental2021} for an example). These inner particles look as if they are inferring mental action policies that minimise their expected free energy, i.e. performing higher order \textit{planning as inference}.

Nested particles satisfying \eqref{eq 1} and \eqref{eq 2} are \emph{comprehensively} metacognitive in the sense that the higher-level internal paths – that is the internal paths of the inner particle – encode posterior beliefs about all [of the sufficient statistics] of the beliefs which the (outer) particle holds about the world. This necessarily entails active metacognitive control over the parameters of lower level beliefs, if the internal blanket comprises active and sensory paths. This can be understood as a form of self-control since the consequence of higher level policies is the manifest dependency of paths internal to the particle.

For information, note that there are also nested particles satisfying \eqref{eq: metacognitive flow} and \eqref{eq 1} with higher-level active and sensory paths -- and thus metacognitive control -- where metacognitive beliefs are only about a subset of lower-level internal paths, making them \emph{partially} metacognitive.

\section{Strange metacognitive particles and the sense of agency}
\label{sec: sense of agency}

Recall that strange particles as defined in \cite{fristonPathIntegralsParticular2023} are such that active paths do not directly influence internal paths, and random fluctuations on the blanket and internal states are negligible.
In \cite{fristonPathIntegralsParticular2023} the authors point out that although a strange particle \eqref{eq: flow} possesses a form of agency -- defined as possessing active paths that depend upon internal paths -- the particle is unable to infer that they are indeed the agent of their actions. Regarding strange particles they write:

\begin{quote}
"From the perspective of someone observing an agent, say a fish, it will look as if the fish searches out particles of food. However, from the point of view of the fish, it believes that it is propelled through water in a fortuitous and benevolent way that delivers food particles to its mouth. In other words, the fish is not aware it is the agent of its actions, it just believes this is how the world works [...]" \cite[p.~27]{fristonPathIntegralsParticular2023}.
\end{quote}

The authors proceed to posit that an agent capable of recognising that their actions are underwritten by agency may require a generative model exhibiting hierarchical depth, namely endowed with beliefs about beliefs. With metacognitive particles as defined here, we can make explicit the justification for this statement. Quite simply, a strange cognitive particle only has beliefs about active and external paths and therefore can never form a belief about the relationship between internal paths $\mu^{(1)}$ and active paths $a^{(1)}$. In order words, they lack the vantage point to realise that they (i.e. their internal paths) are causally upstream of their actions.

A strange, active metacognitive particle, however, has (metacognitive) beliefs about its lower-level internal paths. Hence they have the capacity to form a belief about the causal dependency of $a^{(1)}$ on $\mu^{(1)}$. We posit that this belief can be understood as the particle's \emph{sense of agency}.

We now make this explicit mathematically. The dynamical equations governing our strange particle are \eqref{eq: strange metacognitive flow}, where we have removed the causal direction $a^{(1)} \to \mu^{(1)}$ and random fluctuations on all \emph{particular} paths, i.e. internal, sensory and active paths\cite{fristonFreeEnergyPrinciple2023a}. This lack of a direct influence $a^{(1)} \to \mu^{(1)}$ implies that lower-level internal paths must infer lower-level active paths via lower-level sensory paths (in addition to external paths). The construction in \cite{fristonPathIntegralsParticular2023} yields the following \textit{strange} beliefs (held by a wide class of such strange particles) 
\begin{equation}
\begin{split}
\mu^{(1)} &\mapsto Q_{\mu^{(1)}}(\eta, a^{(1)})\\
     Q_{\boldsymbol{\mu}^{(1)}}(\eta, a^{(1)}) & \triangleq P( \eta, a^{(1)} \mid s^{(1)}).
\end{split}
\end{equation}
Marginalising these beliefs over external paths yields a low level belief about active paths:
\begin{equation}
\begin{split}
 Q_{\mu^{(1)}}(a^{(1)})&\triangleq\int  Q_{\mu^{(1)}}(\eta, a^{(1)}) \d \eta \\
Q_{\boldsymbol{\mu}^{(1)}}(a^{(1)}) & = P(a^{(1)} \mid s^{(1)}) 
\end{split}
\end{equation}

Invoking the comprehensive metacognitive beliefs held by the particle \eqref{eq 2}, the sense of agency is the joint probability distribution
\begin{equation}
\label{eq: agency belief}
\begin{split}
    Q_{\mu^{(2)}}(\mu^{(1)},a^{(1)})&\triangleq Q_{\mu^{(2)}}(\mu^{(1)})Q_{\mu^{(1)}}(a^{(1)})\\
    Q_{\boldsymbol{\mu}^{(2)}}(\mu^{(1)},a^{(1)})&= P(\mu^{(1)} \mid s^{(2)}, a^{(2)})Q_{\mu^{(1)}}(a^{(1)})\\
    Q_{\boldsymbol{\mu}^{(2)}}(\boldsymbol{\mu}^{(1)},a^{(1)})&= P(\mu^{(1)} \mid s^{(2)}, a^{(2)})P(a^{(1)} \mid s^{(1)})\approx P(a^{(1)}, \mu^{(1)} \mid s^{(2)}, a^{(2)})
\end{split}
\end{equation} 
\eqref{eq: agency belief} defines a joint probability distribution over lower level internal and active paths which means that it captures the statistical relationship between (lower level) internal and active paths; hence a sense of agency. A sense of \textit{no agency} would be when we believe that (lower level) internal and active paths are independent so that they do not influence each other:
\begin{equation}
\begin{split}
       Q_{\mu^{(2)}}(\mu^{(1)},a^{(1)})&=Q_{\mu^{(2)}}(\mu^{(1)})Q_{\mu^{(2)}}(a^{(1)})\\
       Q_{\mu^{(2)}}(a^{(1)})&\triangleq \int Q_{\mu^{(2)}}(\mu^{(1)},a^{(1)}) \d \mu^{(1)}.
\end{split}
\end{equation}
Under this perspective, the sense of agency implies a (subjective) measure of the strength of agency as the mutual information between (lower level) internal and active paths under our beliefs:
\begin{equation}
\label{eq: agency KL}
    \dkl\left[Q_{\mu^{(2)}}(\mu^{(1)},a^{(1)}) \mid Q_{\mu^{(2)}}(\mu^{(1)})Q_{\mu^{(2)}}(a^{(1)})\right].
\end{equation}

This measure can be read as a metacognitive framing of empowerment in active inference: c.f., \cite{hafnerActionPerceptionDivergence2020, klyubinEmpowermentUniversalAgentCentric2005}.

\section{Higher forms of metacognition}

\begin{figure}[h]
\centering\includegraphics[width=\textwidth]{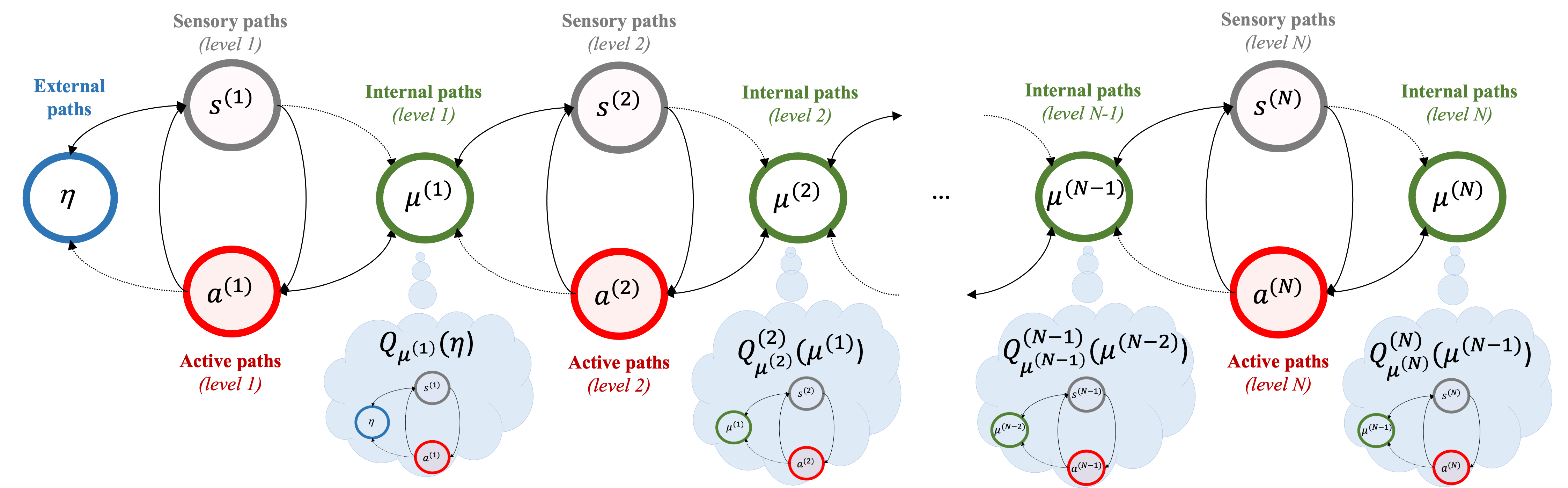}
\caption{\textbf{Multiply nested particles.} This diagram recapitulates the architecture shown in Figure \ref{fig: nested particle}, now extended to $N$ levels of nested metacognition. Each internal path parameterises beliefs about the paths external to it. Note that the inner most paths, $\mu^{(N)}$, are not the target of further metacognitive beliefs.}
\label{fig: multiple nested particle}
\end{figure}

One can go further and consider multiply nested particles, that is, a particle within a particle within a particle, etc. See Figure \ref{fig: multiple nested particle}. Proceeding exactly as above, the internal paths at the highest level can encode beliefs about the internal paths of the level below and so forth. And the internal paths at the lowest level encode beliefs about external paths. In this case we have a particle with as many levels of metacognition as there are nested particles. At each level, the active path can influence the internal path at the level below, which itself influences the sensory path at the current level, which in turn influences the internal path, closing the (mental) action (mental) perception loop. The nested specification is equivalent to saying that the generative model $P(x)$ describing the system is a hierarchical model with as many levels as there are nested components. Finally, multiple nested particles may be $N$-times comprehensively metacognitive in the sense that at each level the internal paths hold beliefs about all of the internal paths at the level below (See Section \ref{sec: cognitive core} Equation \ref{eq: core beliefs}).

Although not depicted here, we can also consider a suitable generalisation of our partially metacognitive particle that exhibits higher order metacognition - as first proposed in \cite{dacostaBayesianMechanicsMetacognitive2024}. Consider the particle of figure \ref{fig: metacognitive particle} with a directed causal chain of internal paths $\mu^{(N)}\rightarrow \mu^{(N-1)} \rightarrow \hdots \rightarrow \mu^{(1)}$ generalising the chain $\mu^{(2)} \rightarrow \mu^{(1)}$. In this case, the internal paths at any level, say $\mu^{(n)}$, can encode beliefs about all of the internal paths at the levels below, i.e. $(\mu^{(n-1)}, \ldots,\mu^{(1)})$. In particular, the internal path at the highest level $\mu^{(N)}$, can encode beliefs about the internal path at the level below $\mu^{(N-1)}$, which itself encodes beliefs about the internal path one level further down $\mu^{(N-2)}$ etc, until $\mu^{(1)}$. In this case, then the particle is $N$-times partially metacognitive, because $\mu^{(N)}$ encodes a chain of $N$ beliefs about $\mu^{(1)}$, and $\mu^{(1)}$ is one of $N$ parameters governing the belief of the particle about the world. 

Note that a multiply nested particle may also have partially metacognitive layers where, say, the $n$-th level internal paths are subdivided into two components $\mu^{(n)}\triangleq(\mu^{(n,1)},\mu^{(n,2)})$ and the internal paths at the level above only hold beliefs about one of those components, e.g. the partially metacognitive belief $Q_{\mu^{(n+1)}}(\mu^{(n,1)})$.

\section{Discussion}

There are many issues that attend this description of metacognitive particles. We take the opportunity to discuss four key points:

\subsection{Infinite regress and the cognitive core}
\label{sec: cognitive core}
In considering multiply nested particles, two natural questions might arise.

The first is whether there is a limitation on the potentially infinite number of nested blankets. The limitation is not given by the treatment in this paper but by the free energy principle \cite{fristonPathIntegralsParticular2023,fristonFreeEnergyPrinciple2023a}. The free energy principle shows that particles will tend to minimise the complexity of their beliefs while maximising their predictive accuracy (about what lies beyond each blanket). Here, we simply say that additional nested parametric depth incurs additional complexity (in terms of additional beliefs), which must be outweighed by the predictive improvement afforded by its inclusion. 
Hence the number of layers remains finite and limited by the complexity-accuracy trade off implicit in free-energy minimization \cite{fristonPathIntegralsParticular2023,fristonFreeEnergyPrinciple2023a}. 

As a result, we conjecture the existence of innermost internal paths $\mu^{(N)}$ that cannot be inferred by higher level metacognitive beliefs (see Figure \ref{fig: multiple nested particle}). This creates a fundamental limitation on self-representation in a system: there will always be a `cognitive core' with internal paths encoding beliefs, whilst never being the target of further higher-order beliefs. This was elegantly expressed by \cite{fristonFreeEnergyMinimizationDarkRoom2012} as: "I can never conceive of what it is like to be me, because that would require the number of recursions I can physically entertain, plus one”. Furthermore, this limitation has been demonstrated by \cite{fieldsPrincipledLimitationsSelfRepresentation2024} within the quantum information theoretic formulation of the free-energy principle, and is in agreement with the notion of an 'irreducible Markov blanket' presented in \cite{ramsteadInnerScreenModel2023}. 

If we subscribe to the notion that subjective experience is related to the information encoded by approximate posterior beliefs, the second question that might arise is how multiple nested Markov blankets can be reconciled with a single, unified, field of experience. This question assumes a relationship between the beliefs parameterised by internal paths and phenomenology. This is an ongoing debate; however, we simply point out that despite the apparent separation between nested layers, the beliefs parameterised by the innermost paths are about the most deeply nested internal paths of all other layers, as well as external paths. Hence the information encoded on the internal Markov blanket captures all the information encoded on lower blankets. We can see this mathematically by expressing the beliefs held by the cognitive core. Assuming one layer of metacognition (i.e. Figure \ref{fig: nested particle}):
\begin{equation}
\begin{split}
Q_{\mu^{(2)}}(\eta,\mu^{(1)}) \triangleq Q_{\mu^{(1)}}(\eta)Q_{\mu^{(2)}}(\mu^{(1)})
\end{split}
\end{equation}
This can be extended to multiply nested particles (i.e. Figure \ref{fig: multiple nested particle}), giving:
\begin{equation}
\label{eq: core beliefs}
\begin{split}
Q_{{\mu}^{(N)}}(\eta,\mu^{(1)},\ldots,\mu^{(N-1)}) \triangleq Q_{{\mu}^{(1)}}(\eta) \prod_{n=2}^{N}
Q_{{\mu}^{(n)}}(\mu^{(n-1)})
\end{split}
\end{equation}

Hence there is a formal object that has the structure of a unified experience if we assume a relationship between phenomenology and information encoded by the approximate posterior belief -- something referred to as extrinsic information geometry \cite{fristonSentienceOriginsConsciousness2020}. 

\subsection{Mathematical characterization}
\label{section:math characterization}

Although we have conceptually described several types of particles in terms of beliefs encoded by different types of internal paths, one important avenue is to determine the exact conditions on their dynamics for these beliefs to exist. These characterisations usually necessitate focusing on a particular functional form for the dynamics driving the system, and sufficient conditions have been given in several cases \cite{fristonPathIntegralsParticular2023,dacostaBayesianMechanicsStationary2021a}. The point for this article thus far is this: these beliefs exist in a wide class of particles subject to the dynamics considered. Metacognitive particles exist – according to our particular definition. Future work will address the specific conditions that particles meeting our definition must meet, which will in turn enable numerical simulations of their belief dynamics.

\subsection{Computational phenomenology}

There is an interesting link between the theoretical treatment presented here and computational models of introspection and meta-awareness describing metacognitive processes such as meditation. As we attempt a mathematical definition of metacognition, this raises the question whether the description presented in this paper could be refined to get closer to our computational and empirical understanding of this phenomenon. Recent advances in active inference modelling of meditation suggest that some aspects can be adequately modelled as active inference under a hierarchical generative model where actions at the higher level modulate the precisions (i.e. inverse covariances) of beliefs encoded by the level below \cite{sandved-smithComputationalPhenomenologyMental2021}. From this perspective, this amounts to the behaviour of multiply nested strange particles \cite{fristonPathIntegralsParticular2023}, where inner particles hold beliefs about active paths that influence the (sufficient statistics of the) beliefs held by the lower-level particle. There is an interesting road ahead where computational models of meditation furnish examples of what we would expect from metacognitive phenomenology, and theoretical treatments along these lines showing general conditions for the phenomenology described by these models to arise. 

For example, a central theme in contemplative practices such Mahamudra or self-inquiry is the first person investigation of the apparent sense of separation from the external world. In these practices, attention is directed towards the impression of perceiving things `over there' from a perspective that is `in here'. Similarly to how the sense of agency can be related to the joint probability distribution over internal and active paths (Section \ref{sec: sense of agency}), this sense of separation finds a potential mathematical candidate in the joint probability distribution over internal and external paths. This belief will necessarily encode the understanding that what is internal to me is separated from what is external, since the Markov blanket enforces a conditional independence of internal and external paths. 

This implies that non-metacognitive particles, in the absence of beliefs about internal paths, might possess a `pre-dual', as well as non-agentive, phenomenology. Duality, understood as a sense of separation, then emerges as a consequence of metacognitive self-representation, i.e. beliefs about beliefs. We propose that further \emph{non-dual} modes of experience might, in principle, be amenable to mathematical description within Bayesian mechanics.

A full exploration of this direction is beyond the scope of this paper, the aim here has been to present an architecture and formalism that might provide a mathematically grounded way of thinking about these fundamental themes in contemplative phenomenology, and metacognitive phenomenology more generally. 

\subsection{Empirical testing}

In principle, it should be possible to experimentally demonstrate whether any given particle (e.g., molecules, mice, or men) falls within the description of particles presented here by examining the structure and nature of their dynamics. Practically, this would require 1) inferring the dynamics from empirical data -- e.g. through Bayesian selection of stochastic dynamical models based on timeseries from a particle and its surrounding milieu -- and then 2) analyzing whether the resulting structure conforms to the mathematical characterization of metacognition discussed above. On a practical level, there are procedures for identifying nested Markov blankets from empirical timeseries, allowing the identification of multiply nested particles. Please see \cite{fristonHierarchicalModelsBrain2008} for a worked example using brain imaging data. Whether these procedures find purchase in decomposing distributed systems across scales, in further empirical studies, remains to be seen.

It is possible to make empirical predictions in regards to the formalisation of the sense of agency. For example, we predict that the KL divergence in (\ref{eq: agency KL}) will correlate with the behaviour, guided by the sense of agency, in the squares task, as participants dynamically close the action-perception loop to determine which square they control \cite{perrykkadEffectUncertaintyPrediction2021}. Interestingly, this may open avenues of research for differences in mental and developmental disorders, such as autism spectrum condition, where metacognition may be implicated.

Another possible avenue of empirical investigation is to examine the nature of the influence of higher level dynamics on lower level paths via (mental) actions mediated by inner active paths. Notice that the mechanism of action between levels, of an active metacognitive particle, is to exert an influence on parameters of lower level beliefs. This has implications for the type of physical interactions that might enable such an influence. For example, in a metacognitive particle we would expect that the dynamics of higher level internal paths be at the origin of processes that impact the dynamics of lower level paths. In neurobiology this leads to predictions, for example, about the existence of neuronal populations whose activity is at the origin of neuromodulatory dynamics. In psychology, this might be read as establishing an attentional set or, more simply, attention \cite{ainleyBodilyPrecisionPredictive2016,duquetteIncreasingOurInsular2017,limanowskiSeeingDarkGrounding2018,sandved-smithComputationalPhenomenologyMental2021,senguptaNeuronalGaugeTheory2016,veissiereThinkingOtherMinds2020}.

\section{Conclusion}
We have proposed a definition of metacognition using statistical physics within the context of Bayesian mechanics and the free-energy principle. Metacognition, understood as the capacity for forming \emph{beliefs about beliefs}, finds a natural mathematical articulation within this formalism. We have further distinguished various types and degrees of metacognition such as strong versus weak, active versus passive. 

In doing so we find a mathematical homologue for the sense of agency (as distinct from the presence of agency in itself) and an argument for why metacognition thus defined is a necessary condition for this sense. We extend the typology of cognitive particles in \cite{fristonPathIntegralsParticular2023} to include particles resembling something closer to ourselves, thus opening a direction for further empirical and mathematical investigation into the physical and informational dynamics at play in complex cognitive organisms.

\appendix

\section{Dynamics of nested particles}
\label{app: flow}

The dynamics of the nested system from Figure \ref{fig: nested particle} is as follows: 

\begin{equation}
\label{eq: metacognitive flow}
     \begin{bmatrix}\dot \eta \\\dot s^{(1)}\\\dot a^{(1)} \\\dot \mu^{(1)}\\\dot s^{(2)}\\\dot a^{(2)} \\\dot \mu^{(2)} \end{bmatrix}(t)=
     \begin{bmatrix}f_\eta(\eta, b^{(1)}) \\ f_{s^{(1)}}(\eta, b^{(1)}) \\ f_{a^{(1)}}(b^{(1)}, \mu^{(1)})\\f_{\mu^{(1)}}(b^{(1)}, \mu^{(1)},b^{(2)})\\ f_{s^{(2)}}(\mu^{(1)}, b^{(2)}) \\ f_{a^{(2)}}(b^{(2)}, \mu^{(2)})\\f_{\mu^{(2)}}(b^{(2)}, \mu^{(2)})\end{bmatrix}(t)+
     \begin{bmatrix}w_\eta \\ w_{s^{(1)}} \\ w_{a^{(1)}} \\ w_{\mu^{(1)}}\\w_{s^{(2)}}\\w_{a^{(2)}}\\w_{\mu^{(2)}}\end{bmatrix}(t),
\end{equation}
where $b^{(i)}=(s^{(i)}, a^{(i)})$ for $i=1,2$ are the lower- and higher- level blankets, respectively. The dynamics governing the strange nested particles discussed in Section \ref{sec: sense of agency} are:
\begin{equation}
\label{eq: strange metacognitive flow}
     \begin{bmatrix}\dot \eta \\\dot s^{(1)}\\\dot a^{(1)} \\\dot \mu^{(1)}\\\dot s^{(2)}\\\dot a^{(2)} \\\dot \mu^{(2)} \end{bmatrix}(t)=
     \begin{bmatrix}f_\eta(\eta, b^{(1)}) \\ f_{s^{(1)}}(\eta, b^{(1)}) \\ f_{a^{(1)}}(b^{(1)}, \mu^{(1)})\\f_{\mu^{(1)}}(s^{(1)}, \mu^{(1)},b^{(2)})\\ f_{s^{(2)}}(\mu^{(1)}, b^{(2)}) \\ f_{a^{(2)}}(b^{(2)}, \mu^{(2)})\\f_{\mu^{(2)}}(b^{(2)}, \mu^{(2)})\end{bmatrix}(t)+
     \begin{bmatrix}w_\eta(t) \\ 0 \\ 0 \\ 0\\0\\0\\0\end{bmatrix},
\end{equation}
so that we removed the dependency of $f_{\mu^{(1)}}$ on $a^{(1)}$ in \eqref{eq: metacognitive flow}.

\printbibliography

@article{perrykkadEffectUncertaintyPrediction2021,
  title = {The Effect of Uncertainty on Prediction Error in the Action Perception Loop},
  author = {Perrykkad, Kelsey and Lawson, Rebecca P. and Jamadar, Sharna and Hohwy, Jakob},
  date = {2021-05-01},
  journaltitle = {Cognition},
  shortjournal = {Cognition},
  volume = {210},
  pages = {104598},
  issn = {0010-0277},
  doi = {10.1016/j.cognition.2021.104598},
  url = {https://www.sciencedirect.com/science/article/pii/S0010027721000172},
  urldate = {2024-06-12}
}

@article{senguptaNeuronalGaugeTheory2016,
  title = {Towards a {{Neuronal Gauge Theory}}},
  author = {Sengupta, Biswa and Tozzi, Arturo and Cooray, Gerald K. and Douglas, Pamela K. and Friston, Karl J.},
  date = {2016-03-08},
  journaltitle = {PLOS Biology},
  shortjournal = {PLOS Biology},
  volume = {14},
  number = {3},
  pages = {e1002400},
  publisher = {Public Library of Science},
  issn = {1545-7885},
  doi = {10.1371/journal.pbio.1002400},
  url = {https://journals.plos.org/plosbiology/article?id=10.1371/journal.pbio.1002400},
  urldate = {2024-05-15},
  langid = {english}
}

@article{veissiereThinkingOtherMinds2020,
  title = {Thinking through Other Minds: {{A}} Variational Approach to Cognition and Culture},
  shorttitle = {Thinking through Other Minds},
  author = {Veissière, Samuel P. L. and Constant, Axel and Ramstead, Maxwell J. D. and Friston, Karl J. and Kirmayer, Laurence J.},
  date = {2020-01},
  journaltitle = {Behavioral and Brain Sciences},
  volume = {43},
  pages = {e90},
  issn = {0140-525X, 1469-1825},
  doi = {10.1017/S0140525X19001213},
  url = {https://www.cambridge.org/core/journals/behavioral-and-brain-sciences/article/abs/thinking-through-other-minds-a-variational-approach-to-cognition-and-culture/9A10399BA85F428D5943DD847092C14A},
  urldate = {2024-05-15},
  langid = {english}
}

@article{ainleyBodilyPrecisionPredictive2016,
  title = {'{{Bodily}} Precision': A Predictive Coding Account of Individual Differences in Interoceptive Accuracy},
  shorttitle = {'{{Bodily}} Precision'},
  author = {Ainley, Vivien and Apps, Matthew A. J. and Fotopoulou, Aikaterini and Tsakiris, Manos},
  date = {2016-11-19},
  journaltitle = {Philosophical Transactions of the Royal Society of London. Series B, Biological Sciences},
  shortjournal = {Philos Trans R Soc Lond B Biol Sci},
  volume = {371},
  number = {1708},
  eprint = {28080962},
  eprinttype = {pmid},
  pages = {20160003},
  issn = {1471-2970},
  doi = {10.1098/rstb.2016.0003},
  langid = {english},
  pmcid = {PMC5062093}
}

@inproceedings{klyubinEmpowermentUniversalAgentCentric2005,
  title = {Empowerment: {{A Universal Agent-Centric Measure}} of {{Control}}},
  shorttitle = {Empowerment},
  booktitle = {2005 {{IEEE Congress}} on {{Evolutionary Computation}}},
  author = {Klyubin, A.S. and Polani, D. and Nehaniv, C.L.},
  date = {2005},
  volume = {1},
  pages = {128--135},
  publisher = {IEEE},
  location = {Edinburgh, Scotland, UK},
  doi = {10.1109/CEC.2005.1554676},
  url = {http://ieeexplore.ieee.org/document/1554676/},
  urldate = {2024-05-15},
  eventtitle = {2005 {{IEEE Congress}} on {{Evolutionary Computation}}},
  isbn = {978-0-7803-9363-9},
  langid = {english}
}

@article{hespDeeplyFeltAffect2021,
  title = {Deeply {{Felt Affect}}: {{The Emergence}} of {{Valence}} in {{Deep Active Inference}}},
  shorttitle = {Deeply {{Felt Affect}}},
  author = {Hesp, Casper and Smith, Ryan and Parr, Thomas and Allen, Micah and Friston, Karl J. and Ramstead, Maxwell J. D.},
  date = {2021-02-01},
  journaltitle = {Neural Computation},
  volume = {33},
  number = {2},
  pages = {398--446},
  issn = {0899-7667, 1530-888X},
  doi = {10.1162/neco_a_01341},
  url = {https://direct.mit.edu/neco/article/33/2/398/95642/Deeply-Felt-Affect-The-Emergence-of-Valence-in},
  urldate = {2022-06-08},
  langid = {english}}

@article{durrOnsagerMachlupFunctionLagrangian1978,
  title = {The {{Onsager-Machlup}} Function as {{Lagrangian}} for the Most Probable Path of a Diffusion Process},
  author = {Dürr, Detlef and Bach, Alexander},
  date = {1978-06-01},
  journaltitle = {Communications in Mathematical Physics},
  shortjournal = {Commun.Math. Phys.},
  volume = {60},
  number = {2},
  pages = {153--170},
  issn = {1432-0916},
  doi = {10.1007/BF01609446},
  url = {https://doi.org/10.1007/BF01609446},
  urldate = {2024-04-11},
  langid = {english}
}

@online{ramsteadInnerScreenModel2023,
  title = {The Inner Screen Model of Consciousness: Applying the Free Energy Principle Directly to the Study of Conscious Experience},
  shorttitle = {The Inner Screen Model of Consciousness},
  author = {Ramstead, Maxwell James and Albarracin, Mahault and Kiefer, Alex and Klein, Brennan and Fields, Chris and Friston, Karl and Safron, Adam},
  date = {2023-04-20T21:21:51},
  doi = {10.31234/osf.io/6afs3},
  url = {https://psyarxiv.com/6afs3/},
  urldate = {2023-05-04},
  langid = {american},
  pubstate = {preprint}
}

@article{fieldsPrincipledLimitationsSelfRepresentation2024,
  title = {Principled {{Limitations}} on {{Self-Representation}} for {{Generic Physical Systems}}},
  author = {Fields, Chris and Glazebrook, James F. and Levin, Michael},
  date = {2024-03},
  journaltitle = {Entropy},
  volume = {26},
  number = {3},
  pages = {194},
  publisher = {Multidisciplinary Digital Publishing Institute},
  issn = {1099-4300},
  doi = {10.3390/e26030194},
  url = {https://www.mdpi.com/1099-4300/26/3/194},
  urldate = {2024-04-08},
  issue = {3},
  langid = {english}
}

@article{sethInteroceptiveInferenceEmotion2013,
  title = {Interoceptive Inference, Emotion, and the Embodied Self},
  author = {Seth, Anil K.},
  date = {2013-11},
  journaltitle = {Trends in Cognitive Sciences},
  shortjournal = {Trends Cogn Sci},
  volume = {17},
  number = {11},
  eprint = {24126130},
  eprinttype = {pmid},
  pages = {565--573},
  issn = {1879-307X},
  doi = {10.1016/j.tics.2013.09.007},
  langid = {english}
}

@article{limanowskiDisAttendingBody2017,
  title = {({{Dis-}})Attending to the Body({{Dis-}})Attending to the Body: {{Action}} and Self-Experience in the Active Inference Framework: {{Action}} and Self-Experience in the Active Inference Framework},
  shorttitle = {({{Dis-}})Attending to the Body({{Dis-}})Attending to the Body},
  author = {Limanowski, Jakub},
  editora = {Metzinger, Thomas K. and Wiese, Wanja and {MIND Group} and {MIND Group}},
  editoratype = {collaborator},
  date = {2017},
  journaltitle = {Philosophy and Predictive Processing},
  publisher = {[object Object]},
  doi = {10.15502/9783958573192},
  url = {http://www.predictive-mind.net/DOI?isbn=9783958573192},
  urldate = {2024-04-04},
  langid = {english}
}

@article{limanowskiSeeingDarkGrounding2018,
  title = {‘{{Seeing}} the {{Dark}}’: {{Grounding Phenomenal Transparency}} and {{Opacity}} in {{Precision Estimation}} for {{Active Inference}}},
  shorttitle = {‘{{Seeing}} the {{Dark}}’},
  author = {Limanowski, Jakub and Friston, Karl},
  date = {2018},
  journaltitle = {Frontiers in Psychology},
  volume = {9},
  issn = {1664-1078},
  url = {https://www.frontiersin.org/articles/10.3389/fpsyg.2018.00643},
  urldate = {2023-06-07}
}

@article{clarkManyFacesPrecision2013,
  title = {The Many Faces of Precision ({{Replies}} to Commentaries on “{{Whatever}} next? {{Neural}} Prediction, Situated Agents, and the Future of Cognitive Science”)},
  shorttitle = {The Many Faces of Precision ({{Replies}} to Commentaries on “{{Whatever}} Next?},
  author = {Clark, Andy},
  date = {2013-05-21},
  journaltitle = {Frontiers in Psychology},
  shortjournal = {Front. Psychol.},
  volume = {4},
  publisher = {Frontiers},
  issn = {1664-1078},
  doi = {10.3389/fpsyg.2013.00270},
  url = {https://www.frontiersin.org/journals/psychology/articles/10.3389/fpsyg.2013.00270/full},
  urldate = {2024-04-04},
  langid = {english}
}

@article{dacostaBayesianMechanicsMetacognitive2024,
  title = {Towards a {{Bayesian}} Mechanics of Metacognitive Particles: {{A}} Commentary on “{{Path}} Integrals, Particular Kinds, and Strange Things” by {{Friston}}, {{Da Costa}}, {{Sakthivadivel}}, {{Heins}}, {{Pavliotis}}, {{Ramstead}}, and {{Parr}}},
  shorttitle = {Towards a {{Bayesian}} Mechanics of Metacognitive Particles},
  author = {Da Costa, Lancelot and Sandved-Smith, Lars},
  date = {2024-03-01},
  journaltitle = {Physics of Life Reviews},
  shortjournal = {Physics of Life Reviews},
  volume = {48},
  pages = {11--13},
  issn = {1571-0645},
  doi = {10.1016/j.plrev.2023.11.014},
  url = {https://www.sciencedirect.com/science/article/pii/S1571064523002002},
  urldate = {2023-12-06}
}

@article{sandved-smithComputationalPhenomenologyMental2021,
  title = {Towards a Computational Phenomenology of Mental Action: Modelling Meta-Awareness and Attentional Control with Deep Parametric Active Inference},
  shorttitle = {Towards a Computational Phenomenology of Mental Action},
  author = {Sandved-Smith, Lars and Hesp, Casper and Mattout, Jérémie and Friston, Karl and Lutz, Antoine and Ramstead, Maxwell J D},
  year = {2021},
  month = jan,
  journal = {Neuroscience of Consciousness},
  shortjournal = {Neuroscience of Consciousness},
  volume = {2021},
  number = {1},
  pages = {niab018},
  issn = {2057-2107},
  doi = {10.1093/nc/niab018},
  url = {https://doi.org/10.1093/nc/niab018},
  urldate = {2022-11-04}
}

@article{dacostaBayesianMechanicsStationary2021a,
  ids = {dacostaBayesianMechanicsStationary2021},
  title = {Bayesian Mechanics for Stationary Processes},
  author = {Da Costa, Lancelot and Friston, Karl and Heins, Conor and Pavliotis, Grigorios A.},
  year = {2021},
  month = dec,
  journal = {Proceedings of the Royal Society A: Mathematical, Physical and Engineering Sciences},
  volume = {477},
  number = {2256},
  eprint = {2106.13830},
  pages = {20210518},
  publisher = {{Royal Society}},
  doi = {10.1098/rspa.2021.0518},
  urldate = {2022-01-05},
  archiveprefix = {arxiv}
}

@article{fristonFreeEnergyMinimizationDarkRoom2012,
  title = {Free-{{Energy Minimization}} and the {{Dark-Room Problem}}},
  author = {Friston, Karl and Thornton, Christopher and Clark, Andy},
  year = {2012},
  month = may,
  journal = {Frontiers in Psychology},
  volume = {3},
  issn = {1664-1078},
  doi = {10.3389/fpsyg.2012.00130},
  urldate = {2020-04-04},
  pmcid = {PMC3347222},
  pmid = {22586414}
}

@article{fristonFreeEnergyPrinciple2019a,
  ids = {fristonFreeEnergyPrinciple2019},
  title = {A Free Energy Principle for a Particular Physics},
  author = {Friston, Karl},
  year = {2019},
  month = jun,
  journal = {arXiv:1906.10184 [q-bio]},
  eprint = {1906.10184},
  primaryclass = {q-bio},
  urldate = {2020-02-29},
  archiveprefix = {arxiv}
}

@article{fristonFreeEnergyPrinciple2023a,
  title = {The Free Energy Principle Made Simpler but Not Too Simple},
  author = {Friston, Karl and Da Costa, Lancelot and Sajid, Noor and Heins, Conor and Ueltzh{\"o}ffer, Kai and Pavliotis, Grigorios A. and Parr, Thomas},
  year = {2023},
  month = jun,
  journal = {Physics Reports},
  series = {The Free Energy Principle Made Simpler but Not Too Simple},
  volume = {1024},
  pages = {1--29},
  issn = {0370-1573},
  doi = {10.1016/j.physrep.2023.07.001},
  urldate = {2023-08-30}
}

@article{fristonHierarchicalModelsBrain2008,
  title = {Hierarchical {{Models}} in the {{Brain}}},
  author = {Friston, Karl},
  editor = {Sporns, Olaf},
  year = {2008},
  month = nov,
  journal = {PLoS Computational Biology},
  volume = {4},
  number = {11},
  pages = {e1000211},
  issn = {1553-7358},
  doi = {10.1371/journal.pcbi.1000211},
  urldate = {2019-08-11},
  langid = {english}
}

@article{fristonPathIntegralsParticular2023,
  title = {Path Integrals, Particular Kinds, and Strange Things},
  author = {Friston, Karl and Da Costa, Lancelot and Sakthivadivel, Dalton A. R. and Heins, Conor and Pavliotis, Grigorios A. and Ramstead, Maxwell and Parr, Thomas},
  year = {2023},
  month = aug,
  journal = {Physics of Life Reviews},
  issn = {1571-0645},
  doi = {10.1016/j.plrev.2023.08.016},
  urldate = {2023-09-04}
}

@article{duquetteIncreasingOurInsular2017,
  title = {Increasing {{Our Insular World View}}: {{Interoception}} and {{Psychopathology}} for {{Psychotherapists}}},
  shorttitle = {Increasing {{Our Insular World View}}},
  author = {Duquette, Patrice},
  date = {2017-03-21},
  journaltitle = {Frontiers in Neuroscience},
  shortjournal = {Front. Neurosci.},
  volume = {11},
  publisher = {Frontiers},
  issn = {1662-453X},
  doi = {10.3389/fnins.2017.00135},
  url = {https://www.frontiersin.org/journals/neuroscience/articles/10.3389/fnins.2017.00135/full},
  urldate = {2024-05-15},
  langid = {english}
}

@article{fristonSentienceOriginsConsciousness2020,
  title = {Sentience and the {{Origins}} of {{Consciousness}}: {{From Cartesian Duality}} to {{Markovian Monism}}},
  shorttitle = {Sentience and the {{Origins}} of {{Consciousness}}},
  author = {Friston, Karl J. and Wiese, Wanja and Hobson, J. Allan},
  year = {2020},
  month = may,
  journal = {Entropy},
  volume = {22},
  number = {5},
  pages = {516},
  publisher = {{Multidisciplinary Digital Publishing Institute}},
  doi = {10.3390/e22050516},
  urldate = {2020-09-11},
  copyright = {http://creativecommons.org/licenses/by/3.0/},
  langid = {english}
}

@article{hafnerActionPerceptionDivergence2020,
  title = {Action and {{Perception}} as {{Divergence Minimization}}},
  author = {Hafner, Danijar and Ortega, Pedro A. and Ba, Jimmy and Parr, Thomas and Friston, Karl and Heess, Nicolas},
  year = {2020},
  month = oct,
  journal = {arXiv:2009.01791 [cs, math, stat]},
  eprint = {2009.01791},
  primaryclass = {cs, math, stat},
  urldate = {2020-11-07},
  archiveprefix = {arxiv}
}

@article{parrMarkovBlanketsInformation2020,
  ids = {parrMarkovBlanketsInformation2019},
  title = {Markov Blankets, Information Geometry and Stochastic Thermodynamics},
  author = {Parr, Thomas and Da Costa, Lancelot and Friston, Karl},
  year = {2020},
  month = feb,
  journal = {Philosophical Transactions of the Royal Society A: Mathematical, Physical and Engineering Sciences},
  volume = {378},
  number = {2164},
  pages = {20190159},
  doi = {10.1098/rsta.2019.0159},
  urldate = {2020-02-29}
}

@book{hohwyThePredictiveMind2013,
  title = {The Predictive Mind},
  author = {Hohwy, Jakob},
  year = {2013},
  month = nov,
  publisher = {{Oxford University Press}},
  doi = {10.1093/acprof:oso/9780199682737.001.0001},
  url = {https://doi.org/10.1093/acprof:oso/9780199682737.001.0001},
  isbn = {978-0-19-968273-7}
}

\end{document}